\documentclass[aps,prl,twocolumn,superscriptaddress,showpacs]{revtex4}
\usepackage{graphicx,epsfig}

\begin{document}

\title
{Quasi-equilibrium during aging of the 2D Edwards-Anderson model}
\author
{S. Franz}
\affiliation
{International Centre for Theoretical Physics, Strada
Costiera 11, P.O. Box 586, I-34100 Trieste, Italy}
\author{V. Lecomte}
\affiliation{International Centre for Theoretical Physics, Strada
Costiera 11, P.O. Box 586, I-34100 Trieste, Italy}
\affiliation{\'Ecole normale sup\'erieure de Cachan, 94235 Cachan,
France}
\author
{R. Mulet}
\affiliation
{International Centre for Theoretical Physics, Strada
Costiera 11, P.O. Box 586, I-34100 Trieste, Italy}
\affiliation
{''Henri Poincare'' Chair of Complex Systems and Superconductivity
Laboratory, Physics Faculty-IMRE, University of Havana, La Habana, Cuba}

\date{\today}

\begin{abstract}
We test the quasi-equilibrium picture of the aging dynamics
-strictly valid in the asymptotic dynamical regime of aging
systems- in the pre-asymptotic aging regime of the two dimensional
Edwards-Anderson spin glass model.  We compare the
fluctuation-dissipation characteristic for spin autocorrelation
function and response with a corresponding one obtained for a
suitably defined new correlation function and its conjugated
response. In agreement with the quasi-equilibrium picture we find
that after a short transient the two corresponding
fluctuation-dissipation ratios (FDR) coincide at equal times.
Moreover we show that, as it happens for the usual FDR, the new
dynamic FDR at finite time coincides with the static one at finite
size.

\end{abstract}


\maketitle

\section{Introduction}

In recent times, following developments in spin glass mean field
theory \cite{CuKu,FrMe}, much emphasis has been put on the study of
off-equilibrium fluctuation-dissipation relations during aging
dynamics in glassy systems. These relations quantify the deviation of
the ratio between correlation functions and conjugated responses from
the one implied by the fluctuation dissipation theorem valid at
equilibrium, and have been posed at the basis of a detailed
thermodynamical and statistical description of the dynamics of glasses
\cite{fravi,biku,selli}. Linear response theory allows to relate
possible asymptotic violation of the fluctuation dissipation theorem to
the failure of ergodicity at the level of the equilibrium measure
\cite{FMPP}.

Given the correlation function $C(t,t_w)$ of a certain observable $A$,
and its conjugated response function $\chi(t,t_w)$ describing the
effect at time $t$ of a field conjugated to $A$ acting from time $0$
to time $t_w$, one can define the fluctuation-dissipation ratio (FDR)
$X(t,t_w)$ from the relation:
\begin{equation}
X(t,t_w)= T\frac{\partial C(t,t_w)/\partial t_w}{\partial
\chi(t,t_w)/\partial t_w}.
\label{eq:FD_rel}
\end{equation}
This is just unity in equilibrium conditions while deviates from it
off-equilibrium and in general depends on
the observable quantity $A$ at hand. In mean field spin glasses the FDR
admits a non trivial limit in the aging regime, where the
correlations assume a scaling invariant behaviour.
Moreover, in the long time limit, taken after the thermodynamic limit,
one can define the function
\begin{equation}
x(q)= \lim_{t, t_w \to \infty} X(t,t_w)|_{C(t,t_w)=q}
\label{eq:x_q}
\end{equation}
which can have a non trivial behavior. When this happens, $x(q)$ turns out
to have an important covariance property under exchange of the given
observable $A$ chosen in the measure of correlation and response. If we
have a correlation function $C_A(t,t_w)$ corresponding to an
observable $A$ and a correlation function $C_B(t,t_w)$ corresponding
to an observable $B$, and we define an auxiliary limiting function
$q_B(q_A)= \lim_{t, t_w \to \infty} C_B(t,t_w)|_{C_A(t,t_w)=q_A}$,
 the following relation holds
\begin{equation}
x_B(q_B(q_A))= x_A(q_A),
\label{eq:x_A}
\end{equation}
the meaning of which is that the functions $X_A(t,t_w)$ and
$X_B(t,t_w)$ coincide asymptotically for equal times.  Moreover, in
ref.~\cite{FMPP} it has been shown that in a large class of finite
dimensional systems with short range interaction $x(q)$ defined in an
out of equilibrium context is deeply related to the nature of the
equilibrium free-energy landscape. In fact, considering the overlap
probability function (OPF) $P(q)$ \cite{MPV} describing the statistics
at equilibrium of the correlations of the observable $A$ in two
configurations chosen with Boltzmann weight, the linear response
theory implies:
\begin{equation}
x(q)= \int_{0}^{q} dq' P(q').
\label{eq:x_q_Pq}
\end{equation}
This equality implies that either $x(q)$ and $P(q)$ are both non
trivial or they are both trivial and could be taken as the starting
point for an experimental measure of the equilibrium OPF from
off-equilibrium dynamics. Of course it does not imply the existence
of some short range system where it is verified non-trivially. Going
through the derivation one realizes that (\ref{eq:x_q_Pq}) expresses
the commutation of the thermodynamic limit and the long time limit as
far as certain susceptibilities are concerned. Notice also that,
thanks to (\ref{eq:x_q_Pq}), equation (\ref{eq:x_A}) expresses the
fact that for two observables $A$ and $B$, couples of states with
identical $q_A$ also have identical $q_B$, i.e. the function
$q_A(q_B)$ defined in the dynamics describes the relation between
different overlaps in equilibrium ergodic components. This property
has been shown to be deeply related to ultrametricity in \cite{FMPP}
where it was found that the combination of relations  (\ref{eq:x_A}) and
(\ref{eq:x_q_Pq}) implies ultrametricity.

The meaning of eq. (\ref{eq:x_q_Pq}) has been clarified in
\cite{fravi} where it has been discussed how $x(q)$ can be related to
the density of metastable states -or quasi-states- with free-energy
density slightly above the minimum, implying that quasi-states of
equal free-energy are selected with equal probability during the
dynamical process. The identity (\ref{eq:x_q_Pq}) and the covariance
property (\ref{eq:x_A}), allow to rationalize \cite{fravi,Fpisa} the
interpretation of the ratios $T/x(q)$ for different values of $q$ as
effective temperatures governing the exchanges of heat among slow
modes evolving on waiting time dependent time scales \cite{cukupe}.

Effective temperatures dependent on $q$ mean that while modes evolving
on the same scale are in equilibrium with each other, heat exchanges
between modes evolving on widely separated scales do not occur. It has
been recently shown \cite{sollich,sollich-2} that in trap models where $x(q)$
has a non trivial $q$ dependence, but ultrametricity does not hold,
different quantities define different FDR. A situation where it would
be difficult to identify the FDR's with effective temperatures.
Conversely, J.-L. Barrat and L. Berthier \cite{JLBB} studied
Lennard-Jones models of glass-forming liquids where an FDR constant in
$q$ seem to describe the off-equilibrium dynamics, and found that
density fluctuations at different wave vectors define the same FDR.

Numerical simulations of 3 and 4 dimensional spin glass
Edwards-Anderson models, comparing extrapolations of the OPF from
finite size systems and extrapolations of the FDR from finite time
indicate the non-triviality -and consistently the identity- of both
functions \cite{mari}. This has been taken as an evidence in favor of
a ``replica symmetry breaking scenario'' (RSB) for finite dimensional
spin glasses.

These extrapolations however have been questioned in a series of
papers showing that the OPF in systems without RSB, i.e where the OPF
is trivial in the thermodynamic limit, can be plagued by severe finite
size effects such that for relatively small systems it appears similar
to what one expect for systems with RSB \cite{bray}. In such
conditions RSB could be wrongly inferred from extrapolations of the
finite size OPF $P(q,L)$ of systems of too small sizes $L$, while the
true $P(q)$ is a trivial single $\delta$-function.  In the same way
one could doubt that off-equilibrium dynamic simulation times are too
short to reliably extrapolate the asymptotic FDR from the finite
$t_w$, and that the true asymptotic one is just a single flat step as in
domain growth problems \cite{barrat}.

On the experimental side it is clear that many systems with slow
aging dynamics are found in pre-asymptotic regimes.  A common
phenomenon is the one of interrupted aging, found e.g. in \cite{melin},
where a slow dynamical regime similar to usual aging eventually
crosses over to equilibrium behavior. In addition, even in three
dimensional spin glasses, the paradigmatic systems where aging
could persist indefinitely, one sees that many quantities are far
from their final values. In particular the experiments of Ocio and
H\'erisson \cite{heriss} where the first experimental
determination of the FDR in spin glasses was achieved, show FD
curves that strongly depend on the waiting time, signifying that
the dynamics is still in some pre-asymptotic regime.  In such
conditions it is of great interest to inquire if the concepts
valid for aging in the asymptotic regime can be adapted to get an
adequate picture of the dynamics on much shorter time regimes.

In this context, one can hypothesize that the identity between static
and dynamic FDR found in 3 and 4 dimension is due to the fact that at
a given time $t_w$ there is a slowly growing length $\xi(t_w)$ over
which the system has effectively equilibrated, and $X(q,t_w)$ would
approximately respect the relation (\ref{eq:x_q_Pq}) with
$P(q,L=\xi(t_w))$.  Such an extension would suggest the approximate
validity at finite time of a quasi-equilibrium picture of the aging
dynamics in which quasi-states with equal free-energy are selected
with equal probability, and the static-dynamic equivalence would just
reflect the properties of the equilibrium landscape of finite size
system.  The hypothesis is rather suggestive as it would provide a
framework to interpret aging properties in an appropriate time scale
even for systems which display interrupted aging.  Here, while in a
certain time window slow evolution and approximated scaling laws for
correlations and/or susceptibilities are observed, the final
asymptotic state is ergodic.  

To test this extension L. Berthier and
A. Barrat \cite{BB} studied the 2D Edwards-Anderson (2DEA) model,
which, on one hand displays strong aging effects at finite times, and
a non trivial OPF for finite size, on the other it is known to finally
reach a paramagnetic state at all finite temperatures.  In that work
it was found that indeed it exists a correspondence $L\to t_w$ such
that the relation (\ref{eq:x_q_Pq}) holds. More recently, Berthier
also studied the three and four dimensional case in a pre-asymptotic
regime obtaining similar results \cite{berthier}.

In the light of the previous considerations about the link among
effective temperatures and time scale separation, these finding appear
rather surprising. Here, no time scale separation is possible, slow
modes have to exchange heat in order to eventually equilibrate.  In
order to save the picture, one can of course hypothesize that this
exchange occur ``adiabatically'', in such a way modes evolving at the
same rate appear able to equilibrate at their effective temperature
with faster or slower modes before exchange heat. If this
consideration applies FDR corresponding to different quantities should
appear approximately equal one to another. In order to test this
hypothesis we consider as in \cite{BB}, the 2D
Edwards-Anderson (2DEA) model where as mentioned aging is interrupted
after a finite relaxation time. In our analysis we define some
suitable correlation and response function not obviously related to
the usual spin autocorrelation and its associated response, and
compare the FDR for both couples of functions. In addition, in two
dimensions we test the equivalence between the static and the dynamic
FDR for the new quantities.

Our results are then compared with analogous measures in the
Viana-Bray diluted spin glass, where (\ref{eq:x_q_Pq}) is known to
hold non-trivially.

The remaining of the paper is organized as follows. In the next
section, we introduce the relevant quantities, then we present and
discuss the results of the simulations and finally the
conclusions are outlined.

\section{\label{sec:mod} Definition of the observables.}

The model we will consider consists in a pair of spin glass systems with
independent random coupling and identical number of spins coupled
through random interactions. Before explicitly introduce the model
let's say a few words to motivate this choice, keeping in mind that
our task will be to compare FDR's corresponding to different
correlation-response couples.  In spin models, defined in terms of an
exchange Hamiltonian $H=\sum_{i<j}^{1,N}J_{ij}S_i S_j$ the natural and
most commonly used choice to probe dynamical correlations is the spin
autocorrelation function at different times: $C(t,t_w)=
N^{-1}\sum_i\langle S_i(t) S_i(t_w)\rangle$
\noindent the corresponding ``zero field cooled'' susceptibility with
respect to small local i.i.d. Gaussian fields $h_i$ with variance
$h_o^2$, introduced in the systems at time $t_w$ and kept on at later
times reads, $\chi(t,t_w) = \frac{1}{ N h_o^2} \sum_i
\overline{\langle h_i S_i(t) \rangle}$ where the over-line denotes the
average over the field.  A second common choice is the ``energy
correlation function'' also known as ``link overlap''
$C_E(t,t_w)=N^{-1}\sum_{i,j}J_{ij}\langle S_i(t) S_j(t_w)\rangle$, and
the associated response: $\chi_E(t,t_w) = \frac{1}{ N h_o^2} \sum_i
\overline{\langle J_{ij} h_j S_i(t) \rangle}$. In mean field, for
Gaussian long range $J_{ij}$'s one can show that in the thermodynamic
limit, choosing the variance of the $J_{ij}$'s to be equal to $1/N$
one has for all times and with no assumption about the dynamics:
\begin{eqnarray}
C_E(t,t_w)&=&C(t,t_w)^2\\ \frac{\partial \chi_E(t,t_w)}{\partial
t_w}&=&C(t,t_w)\frac{\partial \chi_E(t,t_w)}{\partial t_w}
\label{q-q2}
\end{eqnarray}
so that, automatically, for all times, the FDR defined with these
quantities coincide with the one defined with the usual correlations
and response. Analogously at equilibrium, one finds that the relation
$q_E(q)=q^2$ holds and $2 q P_E(q_E(q))=P(q)$ independently of the
ultrametric nature of the organization of the states. 

Then, in order to test
the quasi-equilibrium picture one needs to compare overlaps
non-trivially related one to the other.
Consider, therefore, two copies of spin glass systems, with identical number of
spins, and identically distributed, but independent quenched disorder
and coupled by a random field $R_i$. The Hamiltonian of this compound
system is defined by:
\begin{equation}
H=\sum_{i,j} J^{1}_{ij} S_i^1 S^1_j + \sum_{i,j} J^{2}_{ij} S^2_i S^2_j
+ \sum_{i} R_i S_i^1 S^2_i
\label{eq:hamiltonian}
\end{equation}
\noindent where $J^{1}_{ij}$ and $J^{2}_{ij}$ represent the quenched
disorder in copies 1 and 2 respectively and are quenched variables
respecting the lattice topology and otherwise taken as i.i.d. from a
Gaussian distribution with mean 0 and variance 1. The variables $R_i$
which couple spins with identical label in the two copies have been chosen
randomly with values $R_i=\pm K$.

The dynamical spin autocorrelation function now reads:
\begin{equation}
C(t,t_w)= (2 N)^{-1}\sum_i\overline{\langle S_i^1(t) S_i^1(t_w)+ S^2_i(t)
    S^2_i(t_w)\rangle},
\label{eq:autocorr}
\end{equation}
and the corresponding response
\begin{equation}
\chi(t,t_w) = \frac{1}{2 N h_o^2} \sum_i \overline{\langle h^{1}_i S_i^1(t) +
h^{2}_i S^2_i(t) \rangle}
\label{eq:susc}
\end{equation}

\noindent As second couple of correlation response pair we consider
the spin cross-correlation function \cite{FMPP},
\begin{equation}
C_{cross}(t,t_w)=
(2 N)^{-1}\sum_i\overline{\langle (S_i^1(t) S^2_i(t_w)+ S^2_i(t)
    S_i^1(t_w)) R_i\rangle}
\label{eq:cross_autocorr}
\end{equation}
\noindent and
\begin{equation}
\chi_{cross}(t,t_w) = \frac{1}{2 N h_o^2 K} \sum_i
\overline{\langle
    (h^{2}_i S_i^1(t)
+ h^{1}_i S^2_i(t)) R_i \rangle}
\label{eq:cross_susc}
\end{equation}
Where the $\langle \dots \rangle$ indicates an average over the
initial conditions and $\overline{ \dots}$ over the disorder.  We will
speak about direct correlation and response respectively for
~(\ref{eq:autocorr}) and ~(\ref{eq:susc}) and cross correlation and
response for ~(\ref{eq:cross_autocorr}) and ~(\ref{eq:cross_susc}).

 An explicit formula for $C_{cross}(t,t_w)$ and
$R_{cross}(t,t_w)=-\frac{\partial \chi_{cross}(t,t_w)}{\partial t_w}$
as functionals of $C(t,t_w)$ and $R(t,t_w)=-\frac{\partial
\chi(t,t_w)}{\partial t_w}$ can be given for small $K$ using linear
response theory:
\begin{eqnarray}
\label{prod}
C_{cross}(t,t_w)=K^2 \beta [ \int_0^{t_w} ds C(t,s)R(t_w,s)\\ \nonumber
+\int_0^{t} ds\; C(t_w,s)R(t,s)\ ]\\ \nonumber
R_{cross}(t,t_w)=K^2 \beta \int_{t_w}^{t} ds\; R(t,s)R(s,t_w),
\end{eqnarray}
which shows that if the cross FDR coincide with the direct one, it is
for non trivial reasons.

\section{Results and Discussion}
\label{sec:res}

We studied the cross-quantities in two different systems of Ising
spins with random quenched disorder, the Edward Anderson model, a
bi-dimensional square lattice of spins of size $N=L \times L$ and the
fixed connectivity version \cite{SherringtonSourlas} of the Viana-Bray
model \cite{VB} where the spins are on a random lattice with fixed
connectivity $c=10$ and size $N$. For both models, we considered two
copies with identical number of spins and independent quenched
disorder coupled by a random field $R_i=\pm K$ as in the previous
section with $K=1/2$. For this large value of $K$ we are out of the
linear response regime that allowed us to derive the explicit form of
the cross-quantities as function of the usual ones, but even if the
relations (\ref{prod}) does not hold, there is no reason to believe
that the relation between $X_{cross}$ and $X$ becomes trivial.

In order to have as a reference results for a system where the picture
sketched in the introduction certainly holds we present first the data
of quick simulations of the Viana-Bray model.

\begin{figure}[!htb]
\includegraphics[width=0.94 \columnwidth]{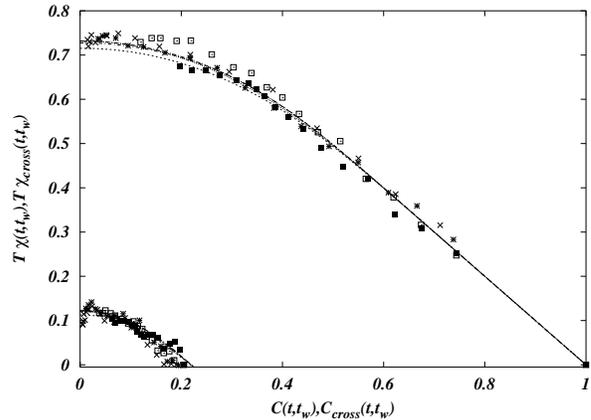}
\caption[0]{Fluctuation dissipation plot for the VB model. A vertical
shift to superimpose the equilibrium curve to the dynamic one. $N=169,
196, 256$ and $324$ and $t_w=10, 10^2, 10^3$ and $10^4$, $T=2.18$ }
\label{fig:fluc-dis-VB}
\end{figure}

In a first test, we compared the FD plots in dynamic simulations of
aging experiments with the static ones obtained through the parallel
tempering technique.  Our results are summarized in figure
\ref{fig:fluc-dis-VB}.  We can observe that, on one hand, the static
characteristics has small finite size dependence, on the other, the
dynamical curves little dependence on the waiting times.  As expected
the dynamic curves coincide with the static ones for the direct
functions. For the cross functions they also coincide provide the
static curves are shifted vertically, this is normal as it should be
noted that, the maximum value of $q_{cross}$ in the equilibrium OPF of
finite systems is unity, while dynamically $C_{cross}(t,t_w)$ is
monotonically decreasing from the value $C_{cross}(t_w,t_w)<1$ for
$t>t_w$. Therefore, one should subtract a constant to the second
integral of $\tilde{P}(q_{cross})$: $S(q_{cross})=\int_{q_{cross}}^1
x_{cross}(q') dq'$ to compare it with the the dynamic function.

We then tested to what extent equation (\ref{eq:x_A}) is valid when
finite systems in statics are compared to systems evolved for finite
aging times in dynamics. From the statics we get the functions $P(q)$ and
$\widetilde{P}(q_{cross})$  where $q$ is defined as the usual (direct)
overlap between two independent replicas ${\bf S}=(S_i^1,S_i^2)$ and
${\bf S'}=(S_i^{1'},S_i^{2'})$ : $q({\bf S},{\bf S'})=\frac{1}{2N}
\sum_{i} S^1_i S_i^{1'}+S^2_i S_i^{2'}$, while $q_{cross}({\bf S},{\bf
S'})=\frac{1}{2N} \sum_{i} R_i(S^1_i S_i^{2'}+S^2_i S_i^{1'})$.
Then using ~(\ref{eq:x_q_Pq}) we derive the
equilibrium quantities $x(q)$ and $\widetilde{x}(q_{cross})$ and
compare them with the results obtained from the dynamic
simulations.

\begin{figure}[!htb]
\includegraphics[width=0.94 \columnwidth]{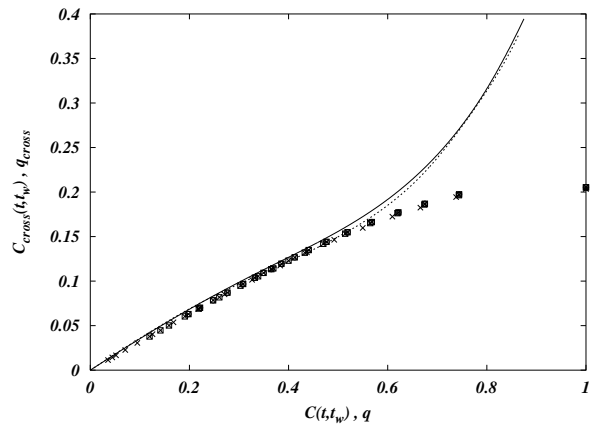}
\caption[0]{Parametric curves of the cross correlation as a function
of the direct one. The lines are the equilibrium curves $N=144$ and $196$
and the points the
dynamic ones $t_w=10, 10^2, 10^3$ and $10^4$.
 We see that the curves approach each other
for values of $q_{cross}$ smaller than $C_{cross}(t_w,t_w)$, which
seems to have reached its $t_w\to\infty$ limit.}
\label{fig:q-vs-qn-VB}
\end{figure}

 Although we did not try to measure the joined probability
$P(q,q_{cross})$, we could extract a function $q_{cross}(q)$ as
implied by the relation (\ref{eq:x_A}) and compare with the one
directly obtained from the dynamics. The results can be seen in
figure \ref{fig:q-vs-qn-VB} where one can see that the static and
the dynamic curves approach each other for values of $q_{cross}$
smaller than $C_{cross}(t_w,t_w)$. This is what one should expect
because , as discussed above, in dynamics this is the largest
value of $C_{cross}(t,t_w)$, and tends to its limit from below for
$t_w \rightarrow \infty$. Conversely in statics, for finite
systems, the probability distribution always extends to values of
$q_{cross}$ larger than the maximum value for an infinite system.

We then pass to the study of the two dimensional system. We
studied the aging dynamics of the 2DEA model with unitary Gaussian
couplings at $T=0.43$, where no
sign of thermalization can be observed in the correlation function up
to waiting times as high as $t_w=10^5$.
In our simulations we used
$h_o^2=0.02$ and checked to be in the linear response regime using the
value $h_o^2=0.01$.
On the equilibrium side, using the parallel tempering technique
\cite{Rieger.et.al} we were able to calculate with good precision, the
spin glass order parameter $P(q)$, as well as the function
$\tilde{P}(q_{cross})$.

\begin{figure}[!htb]
\includegraphics[width=0.94 \columnwidth]{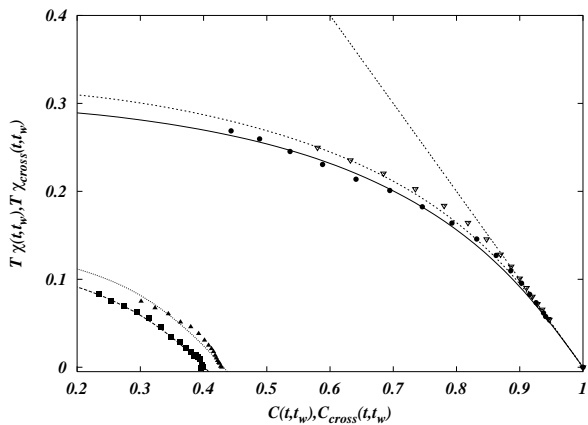}
\caption[0]{Fluctuation dissipation plot for the EA model at
temperature $T=0.43$ and waiting times $t_w=10^2, 10^3$ in comparison with
equilibrium functions for systems of size $L=8,10$. A vertical and an
horizontal shift are necessary to superimpose the equilibrium and the
dynamic curves.}
\label{fig:fluc-dis-EA}
\end{figure}

In figure~\ref{fig:fluc-dis-EA} we present a fluctuation-dissipation
plot for the model. As observed by L. Berthier and
A. Barrat~\cite{BB}, one can superimpose the finite time curves of the
direct functions with the second integral of the equilibrium OPF of
suitable size systems at the same temperature.  The points represent
the data obtained by the dynamic simulation, while the lines those
obtained studying the static of the model using the parallel tempering
technique.  We present for clarity only plots at two different waiting
times, $t_w=10^2$ and $t_w=10^3$, for the dynamic simulations and two
lattice sizes, $L=8$ and $L=10$ for the static ones. The upper curves
correspond to the usual functions (similar curves were already
presented in ref.~\cite{BB}) , while the lower ones, reflects the
cross functions.  Unfortunately, it
turns out that in order to superimpose the static and dynamic curves,
a vertical shift in $S(q)$ is not enough and an horizontal shift,
should also be performed.

In order to understand this point we notice that differently to
what happens for the direct function for which by construction
$C(t,t)=1$, the value of $C_{cross}(t,t)$ evolves in time. In
such conditions the definition of the FDR in terms of the simple
correlation is not necessarily the most appropriate. In fact, the
study of running away systems (e.g. brownian motion or particles
in non confining random potentials \cite{FrMe,cudou}) shows that a
better definition is obtained considering the following
combination of the correlation function
$B_{cross}(t,t_w)=\frac{1}{2}[ C_{cross}(t,t)+
C_{cross}(t_w,t_w)-2 C_{cross}(t,t_w)]$\footnote{We are indebted
with L. Berthier for pointing us out the source of this problem.}.
This obviously is not the only combination which could be used;
e.g. in \cite{sollich-2} it was suggested the use of
$F_{cross}(t,t_w)=C_{cross}(t,t)-C_{cross}(t,t_w)$, however, we
preferred to present the data using  $B_{cross}$, which is
symmetric in the configurations at the two times and admits the
interpretation of a distance between the configurations at times
$t_w$ and $t$. We checked all the results below using $F_{cross}$
instead of $B_{cross}$ which turn out to be equivalent within
our numerical accuracy.

\begin{figure}[!htb]
\includegraphics[width=0.94 \columnwidth]{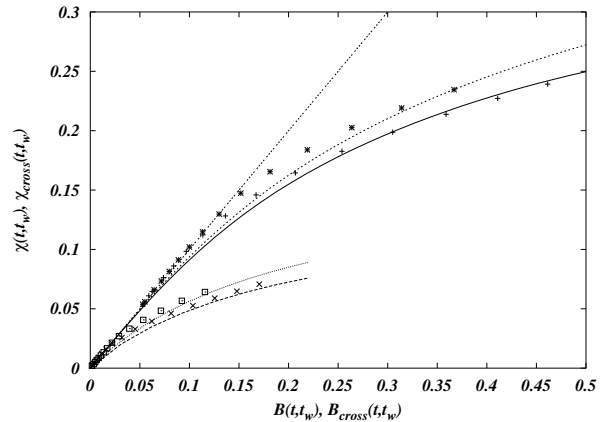}
\caption[0]{Fluctuation dissipation plot for the 2D EA model using $B$
  and $B_{cross}$ as abscissas. The agreement of the dynamical
  characteristic and the static one for the cross quantity is
  comparable to the direct one. In this case no shift of the curves is
  needed.}
\label{statdyn2d.eps}
\end{figure}

\begin{figure}[!htb]
\includegraphics[width=0.94 \columnwidth]{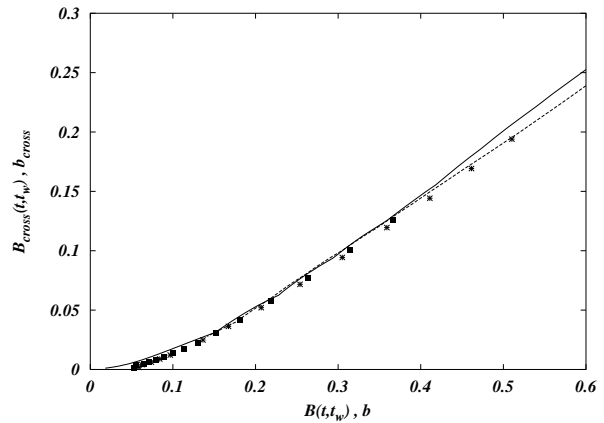}
\caption[0]{Comparison between the relation between $B_{cross}$
and $B$ as obtained directly from the dynamics and supposing eq.
(\ref{eq:x_A}) in
  statics. Solid lines:  the static data for $L=8,10$.
Points: the dynamical data are
  $t_w=10^2,10^3$. The sizes of the static data are $L=8,10$.}
\label{statdyn2d-bis}
\end{figure}

The static analogous of the function $B_{cross}(t,t_w)$ is, given
two configurations ${\bf S}$ and ${\bf S'}$, the quantity
$b_{cross}({\bf S},{\bf S'})=\frac{1}{2}[q_{cross}({\bf S},{\bf
S})+q_{cross}({\bf S'},{\bf S'})-2q_{cross}({\bf S},{\bf S'})]$.
In order to compare the dynamical FDR with the static one we
should be cautious of the symmetry of the Hamiltonian under
contemporary reversal of all the spins. As discussed in
\cite{FMPP} the proper static probability distribution to compare
with the dynamics is not the full distribution but the
distribution modulo the symmetry of the Hamiltonian. For the
function $P(q)$, symmetric under $q\to -q$ one just needs to
consider the positive $q$ part of the function multiplied by 2
(for normalization). In order to eliminate this symmetry in the
distribution $\widetilde{P}(b_{cross})$ in an analogous way, we
can just consider the histogram of the $b_{cross}$ corresponding
to configurations ${\bf S}$ and ${\bf S'}$ such that
$q_{cross}({\bf S},{\bf S'})$ is positive. The results are
presented in figure \ref{statdyn2d.eps} which shows that the
agreement between the static and dynamic curves for the cross
quantities is comparable to the direct ones, for the same lattice
sizes. 

Next we compare the relation between $B_{cross}(t,t_w)$ and
$B(t,t_w)$  obtained directly in dynamics and relating the
values with equal $x$ in statics as explained above. This is shown
in figure \ref{statdyn2d-bis} where we show that for the times and
lengths considered there is a good correspondence between statics
and dynamics.

\begin{figure}[!htb]
\includegraphics[width=0.94 \columnwidth]{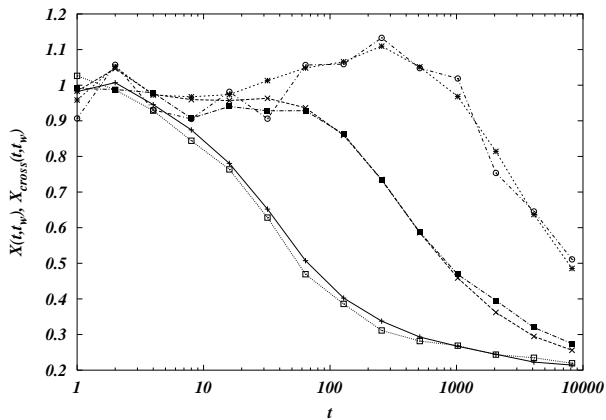}
\caption[0]{Direct comparison of the FDR for the cross and direct
  quantities in the 2D EA model. The temperature is $T=0.43$ and the
  waiting times are, from bottom to top $t_w=10^2,10^3,10^4$. After a
  short transient both quantities do coincide. The glass width is L=512
  and averages are made upon 190 samples.}
\label{fdr2d.eps}
\end{figure}

Finally, given the good quality of our susceptibility data, we
could take the derivatives of the $\chi$ versus $B$
characteristics so to compare directly at equal times the cross
FDR $X_{cross}(t,t_w)$ with the direct one $X(t,t_w)$. The
comparison is shown in figure \ref{fdr2d.eps} which shows that
even for waiting times as short as $t_w=10^2$ the two quantities
are very close to each other. Preliminary results indicate that
this is also the case in three and four dimension.

\section{Conclusions}
\label{sec:conc}

In this paper we have proposed the use of cross correlation functions
and response in disordered systems to probe the existence of effective
temperatures during aging of disordered systems. We have compared the
behaviour of a mean-field model with the one of a paramagnet, where
aging is a transient behavior. We find that the at equal time the FDR
for direct and cross quantities coincide within numerical error after
a short transient. The correspondence among dynamical FDR at finite
time and static one for finite size is confirmed as far as the cross
quantities are concerned. These two findings support on one hand the
idea that aging dynamics can be described in terms of effective
temperatures, on the other that these temperatures are related to the
density of states of finite systems on a scale $L(t_w)$.
Preliminary results in three and four dimension indicate that the same
kind of behaviour is found in these systems for time scales much
shorter than the ones needed to reach an asymptotic state.

\section*{Acknowledgments}

We thank A. Barrat and L. Berthier for many useful discussions
and suggestions. V.L. wishes to thank the ICTP members for kind
hospitality while part of this work was elaborated.

\end{document}